\documentclass[english]{article}
\usepackage[T1]{fontenc}
\usepackage[latin1]{inputenc}
\usepackage{amsmath}
\usepackage{setspace}
\usepackage{amssymb}

\makeatletter
\newcommand{\lyxaddress}[1]{
\par {\raggedright #1
\vspace{1.4em}
\noindent\par}
}

\usepackage{babel}
\makeatother
\begin{document}

\title{Generalized Electromagnetic fields in Chiral Medium}

\author{P. S. Bisht$^{\text{1}}$, Jivan Singh$^{\text{2}}$ and O. P. S.
Negi$^{\text{1}}$}

\maketitle
\begin{singlespace}

\lyxaddress{\begin{center}$^{\text{1}}$Department of Physics\\
 Kumaun University\\
 Soban Singh Jeena Campus\\
 Almora - 263601 (Uttarakhand)\\
 INDIA \par\end{center}}

\lyxaddress{\begin{center}$^{\text{2}}$Department of Physics\\
Govt. Post Graduate College\\
Pithoragarh (Uttarakhand)\\
INDIA\par\end{center}}

\lyxaddress{\begin{center}Email:- ps\_bisht123@rediffmail.com\\
 jgaria@indiatimes.com\\
 ops\_negi@yahoo.co.in\par\end{center}}
\end{singlespace}

\begin{abstract}
The time dependent Dirac-Maxwell's Equations in presence of electric
and magnetic sources are written in chiral media and the solutions
for the classical problem are obtained in unique simple and consistent
manner. The quaternion reformulation of generalized electromagnetic
fields in chiral media has also been developed in compact, simple
and consistent manner.
\end{abstract}

\section{Introduction}

~~~~~Physicists were fascinated about magnetic monopoles since
its ingenious idea was put forward by Dirac \cite{key-1} and also
by Saha \cite{key-2}. So many attempts \cite{key-3,key-4} were made
for the experimental verification of conclusive existence of magnetic
monopoles and after the failure of attempts, the literature \cite{key-5,key-6,key-7}
turned partially negative casting doubts on the existence of such
particles. The work of the Schwinger \cite{key-8} was the first exception
to the argument against the existence of monopoles. At the same time
so many paradoxes were related to the theory of pure Abelian monopoles,
as Dirac's veto, wrong spin-statistics connection \cite{key-9} and
many others \cite{key-10,key-11}. Several problems were soon resolved
by the invention of dyons \cite{key-12,key-13,key-14,key-15} particles
carrying simultaneous existence of electric and magnetic charges.
Fresh interest in this subject was enhanced by the idea given by t'Hooft
\cite{key-16} and Polyakov \cite{key-17} showing that monopoles
are the intrinsic parts of grand unified theories. The Dirac monopole
is an elementary particle but the t'Hooft-Polyakov monopole \cite{key-16,key-17}
is a complicated extended object having a definite mass and finite
size inside of which massive fields play a role in providing a smooth
structure and outside it they vanish rapidly leaving the field configuration
identical to Abelian Dirac monopole. Julia and Zee \cite{key-18}
have extended the idea of t' Hooft \cite{key-16} and Polyakov \cite{key-17}
to construct the classical solutions for non-Abelian dyon. On the
other hand, Prasad and Sommerfield \cite{key-19,key-20} have derived
the analytic stable solutions for the non-Abelian monopoles and dyons
of finite mass by keeping the symmetry of vacuum broken but letting
the self-interaction of Higgs field approaching zero. Such solutions,
satisfying Bogomolnys condition \cite{key-21} are described as Bogomolny-Prasad-Sommerfield
(BPS) monopoles. Kranchenko and coauthors \cite{key-22,key-23} discussed
the Maxwell's equations in homogenous media and developed \cite{key-24}
the quaternionic reformulation of the time-dependent Maxwell's equations
along with the classical solution of a moving source i.e. electron.
They have also demonstrated \cite{key-25} the elctromagnetic fields
in chiral media along with their quaternionic reformulation in a simple
and consistent manner. Recently, we have extended the work of Kranchenko
\cite{key-22,key-23} and reformulated \cite{key-26} the Maxwell's
-Dirac equation in homogenous (isotropic) medium and their quaternionic
forms in a unique and consistent way \cite{key-27}. We have also
described \cite{key-28} the time-harmonic Maxwell's equations for
generalized fields of dyons in a consistent manner. Keeping these
facts in mind, in this paper, we have derived the generalized theory
of Maxwell's -Dirac equation in presence of electric and magnetic
charges in chiral and homogenous (isotropic) medium. The quaternion
analysis of these time dependent Maxwell's equations in presence of
electric and magnetic charges are obtained in unique, simpler and
consistent manner. It has been shown that the quantum equations and
equation of motion represent the dynamics of electric charge similar
to the theory described by Kranchenko \cite{key-22,key-23} in the
absence of magnetic monopole or vice versa.

\section{Generalized Maxwell's Equation of Dyons in Isotropic Medium}

Considering the existence of magnetic monopoles, Dirac \cite{key-1}
generalized the Maxwell's field equations in homogenous (isotropic)
medium in the following manner \cite{key-26};

\begin{eqnarray}
\overrightarrow{\nabla}.\overrightarrow{E} & = & \frac{\rho_{e}}{\epsilon}\nonumber \\
\overrightarrow{\nabla}.\overrightarrow{B} & = & \mu\rho_{m}\nonumber \\
\overrightarrow{\nabla}\times\overrightarrow{E} & = & -\frac{\partial\overrightarrow{B}}{\partial t}-\frac{\overrightarrow{j_{m}}}{\epsilon}\nonumber \\
\overrightarrow{\nabla}\times\overrightarrow{B} & = & \frac{1}{v^{2}}\frac{\partial\overrightarrow{E}}{\partial t}+\mu\overrightarrow{j_{e}}\label{eq:1}\end{eqnarray}
where $\rho_{e}$ and $\rho_{m}$ are respectively the electric and
magnetic charge densities while $\overrightarrow{j_{e}}$ and $\overrightarrow{j_{m}}$
are the corresponding current densities, $\overrightarrow{D}$ is
electric induction vector, $\overrightarrow{E}$ is electric field,
$\overrightarrow{B}$ is magnetic field and$\epsilon_{r}$ and $\mu_{r}$
are defined respectively as relative permitivity and permeability
in electric and magnetic fields.

The electric and magnetic fields of dyons are expresed in homogenous
medium in terms of two potentials as,

\begin{eqnarray}
\overrightarrow{E} & = & -\overrightarrow{\nabla}\phi_{e}-\frac{\partial\overrightarrow{C}}{\partial t}-\overrightarrow{\nabla}\times\overrightarrow{D}\label{eq:2}\\
\overrightarrow{B} & = & -\overrightarrow{\nabla}\phi_{m}-\frac{1}{v^{2}}\frac{\partial\overrightarrow{D}}{\partial t}+\overrightarrow{\nabla}\times\overrightarrow{C}\label{eq:3}\end{eqnarray}
where $\{ C^{\mu}\}=\{\phi_{e},v\,\overrightarrow{C}\}$ and $\{ D^{\mu}\}=\{ v\phi_{m},\overrightarrow{D}\}$
are the two four-potentials associated with electric and magnetic
charges. Let us define the complex vector field $\overrightarrow{\psi}$
in the following form

\begin{eqnarray}
\overrightarrow{\psi} & = & \overrightarrow{E}-i\, v\overrightarrow{\, B}.\label{eq:4}\end{eqnarray}
Equations (\ref{eq:2},\ref{eq:3}) and (\ref{eq:4}) establish the
following relation between generalized field vector and the components
of generalized four-potential as,

\begin{eqnarray}
\overrightarrow{\psi} & = & -\frac{\partial\overrightarrow{V}}{\partial t}-\overrightarrow{\nabla}\phi-iv(\overrightarrow{\nabla}\times\overrightarrow{V)}\label{eq:5}\end{eqnarray}
where $\{ V_{\mu}\}$ is the generalised four-potential of dyons in
homogenous medium and is defined as

\begin{eqnarray}
V_{\mu} & = & \{\phi,\overrightarrow{V}\}\label{eq:6}\end{eqnarray}
i.e. \begin{eqnarray}
\phi & = & \phi_{e}-iv\phi_{m}\label{eq:7}\end{eqnarray}
 and 

\begin{eqnarray}
\overrightarrow{V} & = & \overrightarrow{C}-i\,\frac{\overrightarrow{D}}{v}.\label{eq:8}\end{eqnarray}
Maxwell's field equation (\ref{eq:1}) may then be written in terms
of generalized field $\overrightarrow{\psi}$as

\begin{eqnarray}
\overrightarrow{\nabla}\cdot\overrightarrow{\psi} & = & \frac{\rho}{\epsilon}\label{eq:9}\\
\overrightarrow{\nabla}\times\overrightarrow{\psi} & = & -iv(\mu\overrightarrow{J}+\frac{1}{v^{2}}\frac{\partial\overrightarrow{\psi}}{\partial t})\label{eq:10}\end{eqnarray}
where $\rho$ and $\overrightarrow{J}$ the generalized charge and
current source densities of dyons in homogenous medium given by

\begin{eqnarray}
\rho & = & \rho_{e}-i\frac{\rho_{m}}{v}\label{eq:11}\\
\overrightarrow{J} & = & \overrightarrow{j_{e}}-iv\overrightarrow{j_{m}}.\label{eq:12}\end{eqnarray}
Using equation (\ref{eq:9}) we introduce a new parameter$\overrightarrow{S}$
(i.e. the field current) as

\begin{eqnarray}
\overrightarrow{S} & =\square\overrightarrow{\psi} & =-\mu\frac{\partial\overrightarrow{j}}{\partial t}-\frac{1}{\epsilon}\overrightarrow{\nabla}\rho-iv\mu(\overrightarrow{\nabla}\times\overrightarrow{j})\label{eq:13}\end{eqnarray}
where $\square$is the D'Alembertian operator.

In terms of components of complex potential the Maxell's -Dirac equations
are written as

\begin{eqnarray}
\square\phi & = & v\mu\rho\label{eq:14}\\
\square\overrightarrow{V} & = & \mu\overrightarrow{J}.\label{eq:15}\end{eqnarray}
We may thus write the tensorial form of generalized Maxwell's -Dirac
equations of dyons in homogenous medium as

\begin{eqnarray}
F_{\mu\nu,\nu} & = & j_{\mu}^{e}\label{eq:16}\\
F_{\mu\nu,\nu}^{d} & = & j_{\mu}^{m}.\label{eq:17}\end{eqnarray}
Defining generalized field tensor of dyon as

\begin{eqnarray}
G_{\mu\nu} & = & F_{\mu\nu}-ivF_{\mu\nu}^{d}\label{eq:22}\end{eqnarray}
which yields the following covariant form of generalized field equation
of dyon in homogenous (isotropic) medium i.e.

\begin{eqnarray}
G_{\mu\nu,\nu} & = & J_{\mu}\label{eq:18}\\
G_{\mu\nu,\nu}^{d} & =-i\, J_{\mu}^{\star} & .\label{eq:19}\end{eqnarray}

\section{Generalized Maxwell's Equation for Homogenous Medium in Quaternionic
Form}

A quaternion is defined as 

\begin{eqnarray}
q & = & q_{0}e_{0}+q_{1}e_{1}+q_{2}e_{2}+q_{3}e_{3}\label{eq:20}\end{eqnarray}
where $q_{0},q_{1},q_{2},q_{3}$are real numbers and called the components
of the quaternion $q$ and the quaternion units $e_{0},e_{1},e_{2},e_{3}$
satisfy the following multiplication rules;

\begin{eqnarray}
e_{0}^{2} & = & 1\nonumber \\
e_{j}e_{k} & = & -\delta_{jk}+\epsilon_{jkl}e_{l}\label{eq:21}\end{eqnarray}
where $\delta_{jk}$ and $\epsilon_{jkl}$ ($j, k, l$= 1,2,3 and
$e_{0}=1$) are respectively the Kronecker delta and three-index Levi-Civita
symbol. For any quaternion, there exists a quaternion conjugate

\begin{eqnarray}
\overline{q} & = & q-q_{1}e_{1}-q_{2}e_{2}-q_{3}e_{3}=q_{0}-\overrightarrow{q}.\label{eq:22}\end{eqnarray}
Quaternion conjugate is an automorphism of ring of quaternion i.e.

\begin{eqnarray}
(\overline{pq}) & = & (\overline{q})(\overline{p}).\label{eq:23}\end{eqnarray}
The norm of a quaternion is given as

\begin{eqnarray}
N(q)=q.\overline{q} & = & \overline{q}.q=q_{0}^{2}+q_{1}^{2}+q_{2}^{2}+q_{3}^{2}=|q|^{2}.\label{eq:24}\end{eqnarray}
The inverse of a quaternion is also a quaternion 

\begin{eqnarray}
q^{-1} & = & \frac{\overline{q}}{|q|^{2}}.\label{eq:25}\end{eqnarray}
The quaternionic form of differential operator may be defined as \cite{key-27}

\begin{eqnarray*}
\boxdot & = & (-\frac{i}{v}\partial_{t}+D)\end{eqnarray*}
and its quaternion conjugate as 

\begin{eqnarray}
\overline{\boxdot} & = & (-\frac{i}{v}\partial_{t}-D)\label{eq:26}\end{eqnarray}
where $D=\partial_{1}e_{1}+\partial_{2}e_{2}+\partial_{3}e_{3}.$Defining
the complex vector wave function of generalized electromagnetic field
as \cite{key-26}

\begin{eqnarray}
\overrightarrow{\psi} & = & \overrightarrow{E}-iv\overrightarrow{B}.\label{eq:27}\end{eqnarray}
We can now express the generalized four-potential, four current, electric
field and magnetic in quaternionic form as \cite{key-27} ;

\begin{eqnarray}
V & = & -i\frac{\phi}{v}+V_{1}e_{1}+V_{2}e_{2}+V_{3}e_{3}\label{eq:28}\end{eqnarray}

\begin{eqnarray}
J & = & -i\rho v+J_{1}e_{1}+J_{2}e_{2}+J_{3}e_{3}\label{eq:29}\end{eqnarray}

\begin{eqnarray}
E & = & E_{1}e_{1}+E_{2}e_{2}+E_{3}e_{3}\label{eq:30}\end{eqnarray}

\begin{eqnarray}
B & = & B_{1}e_{1}+B_{2}e_{2}+B_{3}e_{3}\label{eq:31}\end{eqnarray}
Operating first equation of (\ref{eq:26}) to equation (\ref{eq:27})
and using equation (\ref{eq:21} ), we get

\begin{eqnarray}
\overline{\boxdot}\psi & = & J.\label{eq:32}\end{eqnarray}
Similarly we operate first equation of (\ref{eq:26}) to equation
(\ref{eq:29}) and using equation (\ref{eq:21} ) , we get

\begin{eqnarray}
\boxdot J & = & S.\label{eq:33}\end{eqnarray}
Equations (\ref{eq:32}) and (\ref{eq:33} ) are the quaternion forms
of field equations associated respectivly with the generalized potential
and currents of dyons in homogeneous isotropic medium. These equations
are invariant under quaternion transformations as well as with homogeneous
Lorentz transformations. Instead of four sets of differential equations
of fileld equations we may write only one set of quaternion differential
equations in compact ,simple and consistent manner. We may reinterpret
our results as that the algebra over the field of real numbers corresponds
four sets of differential equations given by equation (\ref{eq:1}
), the algebra over the field of complex numbers corresponds to two
sets of differential equations given by equation (\ref{eq:9}-\ref{eq:10}
) while the algebra over the field of quaternion variables corresponds
only to one set of differential equations given by equation (\ref{eq:32}
). As such equations (\ref{eq:32}) and (\ref{eq:33} ) are the quaternion
forms of differential equations ( \ref{eq:9}-\ref{eq:10}and \ref{eq:13}
) and may be visualzed as the quaternion reformulation of generalized
potential and current of dyons in isotropic homogeneous medium in
compact, simpler and cosistent manner.

\section{Generalized Electromagnetic Fields in Chiral Media}

Let us consider the generalized Maxwell's equations for dyons in homogenous
(isotropic) medium as

\begin{eqnarray}
\overrightarrow{\bigtriangledown}.\widetilde{E}(x) & = & \frac{\rho_{e}(x)}{\epsilon}\nonumber \\
\overrightarrow{\bigtriangledown}.\widetilde{H}(x) & = & \rho_{m}(x)\nonumber \\
\overrightarrow{\bigtriangledown}\times\widetilde{E}(x) & = & -\frac{\widetilde{j_{m}}(x)}{\epsilon}+i\omega\widetilde{B}(x)\nonumber \\
\overrightarrow{\bigtriangledown}\times\widetilde{H}(x) & = & \mu\widetilde{j_{e}}(x)-i\omega\widetilde{D}(x).\label{eq:34}\end{eqnarray}
In the above equation we have taken the electric field $\overrightarrow{E}$
and magnetic field $\overrightarrow{B}$ is time harmonic. Considering
electric and magnetic field as under \cite{key-28}

\begin{eqnarray}
\overrightarrow{E}(x,t) & = & Re(\overrightarrow{E}(x)e^{-i\omega t}\nonumber \\
\overrightarrow{B}(x,t) & = & Re(\overrightarrow{B}(x)e^{-i\omega t}.\label{eq:35}\end{eqnarray}
Here we consider the Born-Fedorov consistutive equations \cite{key-29,key-30,key-31,key-32},

\begin{eqnarray}
\widetilde{D}(x) & = & \epsilon(\widetilde{E}(x)+\beta\bigtriangledown\times\widetilde{E}(x))\nonumber \\
\widetilde{B}(x) & = & \mu(\widetilde{H}(x)+\beta\bigtriangledown\times\widetilde{H}(x))\label{eq:36}\end{eqnarray}
where $\epsilon,\mu,$ and $\beta$ are premittivity, permeability
and chiral parameter respectively. If the medium is isotropic, the
cartesian field components are given by 

\begin{eqnarray}
\widetilde{D_{x}} & = & \varepsilon\widetilde{E_{x}}+\varepsilon\beta(\frac{\partial\widetilde{E_{z}}}{\partial y}-\frac{\partial\widetilde{E_{y}}}{\partial z})\nonumber \\
\widetilde{D_{y}} & = & \varepsilon\widetilde{E_{y}}+\varepsilon\beta(\frac{\partial\widetilde{E_{x}}}{\partial z}-\frac{\partial\widetilde{E_{z}}}{\partial x})\nonumber \\
\widetilde{D_{z}} & = & \varepsilon\widetilde{E_{z}}+\varepsilon\beta(\frac{\partial\widetilde{E_{y}}}{\partial x}-\frac{\partial\widetilde{E_{x}}}{\partial y})\nonumber \\
\widetilde{B_{x}} & = & \mu\widetilde{B_{x}}+\mu\beta(\frac{\partial\widetilde{B_{z}}}{\partial y}-\frac{\partial\widetilde{B_{y}}}{\partial z})\nonumber \\
\widetilde{B_{y}} & = & \mu\widetilde{B_{y}}+\mu\beta(\frac{\partial\widetilde{B_{x}}}{\partial z}-\frac{\partial\widetilde{B_{z}}}{\partial x})\nonumber \\
\widetilde{B_{z}} & = & \mu\widetilde{B_{z}}+\mu\beta(\frac{\partial\widetilde{B_{y}}}{\partial x}-\frac{\partial\widetilde{B_{x}}}{\partial y}).\label{eq:37}\end{eqnarray}
Applying the condition given by (\ref{eq:36}) into third and fourth
equation of equation (\ref{eq:35}) we get,

\begin{eqnarray}
\nabla\times\widetilde{E}(x) & = & -\frac{\widetilde{j_{m}}(x)}{\varepsilon}+i\omega\mu(\widetilde{H}(x)+\beta\nabla\times\widetilde{H}(x))\nonumber \\
\nabla\times\widetilde{H}(x) & = & -\widetilde{j_{e}}(x)-i\omega\varepsilon(\widetilde{E}(x)+\beta\nabla\times\widetilde{E}(x)).\label{eq:38}\end{eqnarray}
If we introduce the following notations

\begin{eqnarray}
\widetilde{E}(x) & = & -\sqrt{\mu}\overrightarrow{E}(x)\nonumber \\
\widetilde{H}(x) & = & \sqrt{\varepsilon}\overrightarrow{H}(x)\nonumber \\
\widetilde{j_{e}}(x) & = & \sqrt{\varepsilon}\overrightarrow{j_{e}}(x)\nonumber \\
\widetilde{j_{m}}(x) & = & \sqrt{\varepsilon}\overrightarrow{j_{m}}(x),\label{eq:39}\end{eqnarray}
 the first and second equation of equation (\ref{eq:35}) and equation
(\ref{eq:38}) reduce to the following differential equations as,

\begin{eqnarray*}
\bigtriangledown.\overrightarrow{E} & =- & \frac{\rho_{e}(x)}{\epsilon\sqrt{\mu}}\\
\bigtriangledown.\overrightarrow{H} & = & \frac{\rho_{m}(x)}{\sqrt{\varepsilon}}\end{eqnarray*}

\begin{eqnarray}
\nabla\times\overrightarrow{E}(x) & = & -\frac{\overrightarrow{j_{m}}}{\sqrt{\mu\varepsilon}}-i\alpha(\overrightarrow{H}(x)+\beta\nabla\times\overrightarrow{H}(x))\nonumber \\
\nabla\times\overrightarrow{H}(x) & = & \overrightarrow{j_{e}}(x)+i\alpha(\overrightarrow{E}(x)+\beta\nabla\times\overrightarrow{E}(x))\label{eq:40}\end{eqnarray}

where $\alpha=\frac{\omega}{v}$ is denoted as the wave number.

\section{Maxwell's Equation for Dyons in Quaternionic Chiral Media}

Let us consider the following purely vectorial biquaternionic functions
as,

\begin{eqnarray}
\overrightarrow{m} & (x)= & \overrightarrow{E}(x)+i\overrightarrow{H}(x)\label{eq:41}\\
\overrightarrow{n} & (x)= & \overrightarrow{E}(x)-i\overrightarrow{H}(x).\label{eq:42}\end{eqnarray}
Operating the quaternionic differential operator $D$ on equation
(\ref{eq:41}), we get

\begin{eqnarray}
D\overrightarrow{m}(x) & = & \frac{\rho(x)}{\varepsilon\sqrt{\mu}}(1+\alpha\beta)-\alpha\overrightarrow{m}(x)-\alpha\beta D\overrightarrow{m}(x)+i\overrightarrow{j}(x).\label{eq:43}\end{eqnarray}
Thus the complex quaternionic function $\overrightarrow{m}(x)$ satisfies
the following equation

\begin{eqnarray}
(D+\frac{\alpha}{(1+\alpha\beta)})\overrightarrow{m}(x) & = & \frac{\rho(x)}{\epsilon\sqrt{\mu}}+i\frac{\overrightarrow{j}(x)}{(1+\alpha\beta)}.\label{eq:44}\end{eqnarray}
By analogy of equation (\ref{eq:44}), we obtain the equation for
$\overrightarrow{n}$ as,

\begin{eqnarray}
(D-\frac{\alpha}{(1-\alpha\beta)})\overrightarrow{n}(x) & = & \frac{\rho(x)}{\epsilon\sqrt{\mu}}-i\frac{\overrightarrow{j}(x)}{(1-\alpha\beta)}.\label{eq:45}\end{eqnarray}
On using equation (\ref{eq:34}) and equation (\ref{eq:40}), the
continuity equation is described in the following manner,

\begin{eqnarray}
div\overrightarrow{j{}_{e}} & = & i\frac{\alpha\rho_{e}}{\varepsilon\sqrt{\mu}}\nonumber \\
div\overrightarrow{j{}_{m}} & = & i\frac{\alpha\rho_{m}}{v\sqrt{\varepsilon}}.\label{eq:46}\end{eqnarray}
Introducing the notations

\begin{eqnarray}
\alpha_{1} & = & \frac{\alpha}{(1+\alpha\beta)}\nonumber \\
\alpha_{2} & = & \frac{\alpha}{(1-\alpha\beta)}\label{eq:47}\end{eqnarray}
and using equations (\ref{eq:46}) and (\ref{eq:47}), we may write
the differential equations (\ref{eq:44} )and (\ref{eq:45} ) in the
folowing forms i.e.,

\begin{eqnarray}
(D+\alpha_{1})\overrightarrow{m}(x) & = & -i\frac{div\overrightarrow{j}(x)}{\alpha}-i\frac{\overrightarrow{j}(x)\alpha_{1}}{\alpha}\nonumber \\
=- & \frac{i}{\alpha} & (\alpha_{1}\overrightarrow{j}(x)+div\overrightarrow{j}(x))\nonumber \\
(D-\alpha_{2})\overrightarrow{n}(x) & = & \frac{i}{\alpha}(\alpha_{1}\overrightarrow{j}(x)-div\overrightarrow{j}(x)).\label{eq:48}\end{eqnarray}
In the above equations if we put $\beta=0$, we get the quaternionic
form of the generalized Maxwell's equation of dyons in the absence
of chiral medium. In general the wave numbers $\alpha_{1}$and $\alpha_{2}$
are different and physically characterize the propogation of waves
of opposing circular polarizations.

\section{Discussion}

In the fore going analysis, the equations (\ref{eq:36}) showing the
dependence of the electric displacement vector and the magnetic induction
vector on the electric and magnetic fields, do not take into account
the chirality of the medium. Instead, they depend upon only on the
conductivity, electric permittivity and magnetic permeability of the
medium. Chirality is described as the asymmetry in the molecular structure
where a molecule is chiral if it cannot be superimposed onto its mirror
image. Presence of chirality results in the rotation of electromagnetic
fields and its observable, particular in the microwave range even
for the case of a particle consisting electric and magnetic charges
(i.e. a dyon). Such experimental observations may be used in physical
chemistry to characterize molecular structure. The present theory
of generalized electrodynamics of dyons leads the connection between
the mechanical parameters with the chirality and dilectric properties
of the brain tissue considered as a bioplasama . Hence the proposal
for dyonic bioplasama is being considered. Our theory reduces to the
theories described earlier \cite{key-22,key-23,key-24,key-25,key-29,key-30,key-31,key-32}
for the case of elecrtic charge in the absence of magnetic mponopole
on dyon and cosequently the theories of pure monopole be described
from duality in the absence of electric charge on dyons.

\end{document}